\documentclass[oneside, twocolumn, 11pt]{article}
\usepackage[utf8]{inputenc}
\usepackage[left=2.0cm, right=2.0cm, top=2.0cm, bottom=2.0cm]{geometry}
\usepackage{fancyhdr}
\usepackage[compact]{titlesec}
\usepackage{inconsolata}
\usepackage{sectsty}
\usepackage[none]{hyphenat}
\usepackage{mathpazo}
\usepackage[medium]{inter}
\allsectionsfont{\bfseries\sffamily}

\usepackage{authblk}
\usepackage{doi}
\usepackage{svg}
\usepackage{orcidlink}
\usepackage{xcolor}
\usepackage{url}
\usepackage{hyperref}
\usepackage[backend=bibtex, style=chem-rsc, sorting=none, doi=true, 
articletitle=true, url=true, maxnames=10]{biblatex}
\usepackage{graphicx}

\definecolor{yardgreen}{HTML}{58D0BB}
\definecolor{yardred}{HTML}{F08080}
\newcommand{\datatractor}{\texttt{Datatractor}\,}
\newcommand{\filetype}{\textcolor{black}{\texttt{FileType}}}
\newcommand{\filetypes}{\textcolor{black}{\texttt{FileType}s}}
\newcommand{\extractor}{\textcolor{black}{\texttt{Extractor}}}
\newcommand{\extractors}{\textcolor{black}{\texttt{Extractor}s}}

\title{\datatractor: Metadata, automation, and registries for extractor interoperability in the chemical and materials sciences}

\author[1,2]{Matthew L. Evans \orcidlink{0000-0002-1182-9098}\thanks{matthew@datalab.industries}}
\author[1,3]{Gian-Marco Rignanese\orcidlink{0000-0002-1422-1205}}
\author[4]{David Elbert \orcidlink{0000-0002-2292-180X}}
\author[5]{Peter Kraus \orcidlink{0000-0002-4359-5003}\thanks{peter.kraus@tu-berlin.de}}

\affil[1]{UCLouvain, Institute of Condensed Matter and Nanosciences (IMCN), Chemin des \`{E}toiles 8, Louvain-la-Neuve 1348, Belgium}
\affil[2]{Matgenix SRL, A6K Advanced Engineering Center, Charleroi, Belgium}
\affil[3]{WEL Research Institute, Avenue Pasteur 6, 1300 Wavre, Belgium}
\affil[4]{Hopkins Extreme Materials Institute (HEMI), Johns Hopkins University, Baltimore, MD, USA\vspace{-0.05in}}
\affil[5]{Technische Universität Berlin, Chair for Advanced Ceramic Materials, Hardenbergstr. 40, 10623 Berlin, Germany}
\date{\today\\[0.2in]\emph{``When tillage begins, other arts will follow."}~---~Daniel Webster}

\begin{document}

\maketitle

\begin{abstract}
    Two key issues hindering the transition towards FAIR data science are the poor discoverability and inconsistent instructions for the use of data extractor tools, i.e., how we go from raw data files created by instruments, to accessible metadata and scientific insight. If the existing format conversion tools are hard to find, install, and use, their reimplementation will lead to a duplication of effort, and an increase in the associated maintenance burden is inevitable. 
    
    The \datatractor framework presented in this work addresses these issues. First, by providing a curated registry of such extractor tools their discoverability will increase. Second, by describing them using a standardised but lightweight schema, their installation and use is machine-actionable. Finally, we provide a reference implementation for such data extraction. The \datatractor framework can be used to provide a public-facing data extraction service, or be incorporated into other research data management tools providing added value.
\end{abstract}

\section{Introduction}

The majority of chemistry and materials science research is digital. Fine-grained, unfettered access to the raw data we produce is therefore a crucial part of doing robust, reproducible science.~\cite{Brinson2024}
Typically, the starting point of any scientific analysis is the extraction of data and metadata from a "raw data" file that was created by a piece of apparatus or a piece of software.
Such raw data files are either analysed in proprietary software provided by a vendor directly, or converted into a more accessible and interoperable format using extractors.

How easy or feasible the extraction process is depends heavily on the characteristics of the produced raw data file. In the ideal case, these raw data files would use generic, structured and self-describing formats (HDF5, NeXuS). However, there are limited incentives for instrument and software vendors to use such formats. In addition to operating in a commercial environment, where "vendor lock-in" is often thought of as a positive,~\cite{Gupta2023} producing data in self-describing formats would require community consensus on an appropriate file format, and especially semantics. Domains where such consensus exists are still rare (e.g., FITS in astronomy,~\cite{Mink2015} NetCDF in climate modelling,~\cite{Eaton2003} or CIF in crystallography~\cite{Hall1991}). 

Instead, we are often left with proprietary, binary, or otherwise obfuscated formats, that can require significant additional work to reverse engineer.
These formats can change unpredictably between versions and have the effect (intentional or otherwise) of locking you into the software ecosystem of the hardware vendor.
This presents a significant barrier, both to the day-to-day scientific workflow and to open science generally, since the raw data cannot be made FAIR. Interoperability is discouraged by design, reproducibility is discouraged as a consequence.~\cite{Landrum2012, Feger2020}

By opening up file format definitions and encouraging sharing and reuse of extractors, scientific workflows can become more modular and interoperable with each other, lowering the barrier to more advanced analysis that may rely on the processing of raw data from multiple techniques simultaneously (e.g., in situ or operando measurements).~\cite{Kraus2022, Senocrate2024}
As an increasing number of laboratories undergo the digital transformation and take more steps towards lab automation, the need for robust, open data exchange and storage (via files or otherwise) grows more pressing.

Many communities have developed high-quality open source packages that can parse or extract data from the raw data file formats in their domain. 
In practice, such extractors often start out as ephemeral software (scripts, or notebooks), written to be used once and discarded. However, once shared with colleagues, such ephemeral software nearly always turns into bespoke software used within a research group, handed down generations of PhD students.~\cite{Fehr2019} Such software only rarely evolves into an open source solution developed and maintained by the broader community. 
Therefore, there is a significant duplication of effort, both in the initial development, and then additionally in the ongoing maintenance and upkeep (see the "research code" vs "production-grade code" distinction of Lehtola~\cite{Lehtola2023a}), as the raw data file formats may change and grow in complexity.
New solutions are also sometimes developed out of ignorance of the existing solutions, and the more niche the file format, the less likely it is that the development of an extractor will continue beyond the ephemeral or bespoke stage.

Additionally, the developers working at the data management platform level may have to (re)implement support for all of the desired formats in their community. Yet, the challenges faced when integrating such data extractors are broadly common between techniques. 
This redundant work could therefore be tackled at the level of the extractors themselves.~\cite{Skluzacek2019} However, achieving such a distributed development model would require a discoverability mechanism for existing, tested, and maintained extractors, in order to avoid reinventing the wheel~\cite{Skluzacek2019,Skluzacek2023}.

Rather than making an attempt to produce yet another centralised package, or to conflict with ongoing important work in developing new flexible data standards that vendors can directly conform to,~\footnote{We wish to highlight the current efforts in the extension of the NeXuS file format to cover the areas of material synthesis and characterisation, i.e. the Area A and Area B work of the FAIRmat project.} here we describe a solution to the problem of the ongoing discoverability and interoperability of \emph{existing} extractors.
In the current work, we present a framework for describing the extractors themselves, their usage, and their inputs (i.e., the "raw data" file formats); description of the extractor output is not part of the current framework and will be a topic of future work.
Our approach stems from discussions within the context of the Materials Research Data Alliance (MaRDA) extractors working group (WG),~\footnote{See \url{https://www.marda-alliance.org/working-group/wg7-automated-metadata-extractors} for description of the WG and \url{https://github.com/marda-alliance/metadata_extractors/discussions} for the discussions.} which culminated in the \datatractor project.

In the rest of this paper, we first describe the design of {\datatractor} and technical implementation details of its three components. 
We also describe how one would announce their extractor within the {\datatractor} project, and how one would use such extractors to extract their raw data. 
Then we outline the downstream (i.e., platform users) and upstream (i.e., extractor developers) use cases which might be enabled by this approach, with a particular focus on the dual use-case of \emph{datalab}~\cite{Evans2024}, a platform that also contributes extractors.
We end with a call-to-arms for users and developers to contribute to the registry and grow this decentralized resource.

\section{Design \& implementation}

The main goal of \datatractor is to provide a place where extractor code can be promoted as FAIR software objects in and of themselves.~\cite{Barker2022}, whilst simultaneously enhancing extractor discoverability and reducing duplication of effort across data platforms and individual researchers.
In order to meet these goals in a scalable and vendor-neutral way, we must go beyond a simple library that bundles all existing extractors with a uniform interface, and instead require persistent schemas for code authors to describe fully the intent and usage of their code in a machine-actionable way.

At the same time, it is crucial that scientists and less technical users feel empowered to use these tools, and indeed, to ask more of their tools. As such, our goal was to provide a simple searchable graphical web interface
and a zero-dependency example implementation that could be easily adopted by individual users and platforms alike, alongside their existing tools.

To this end, \datatractor has three main components:
\begin{enumerate}
\item \texttt{schemas}: semantic schemas for file types and extractor codes, their definitions, installation and usage instructions, 
\item \texttt{yard}: a registry of community-sourced metadata, describing file types and extractor codes, which is machine-actionable and user-searchable,
\item \texttt{beam}: a reference implementation that can parse files according to these declarative definitions present in the \texttt{yard} and elsewhere.
\end{enumerate}

\subsection{Semantic schemas}

Two schemas have been designed, one for file types (\filetype) and one for extractor codes (\extractor). Their design was borne out of discussions in the MaRDA WG and then iteratively co-developed alongside the registry and reference implementation (\texttt{yard} and \texttt{beam}, respectively).

\begin{figure}[tb!]
    \centering
    \includegraphics[width=1.0\linewidth]{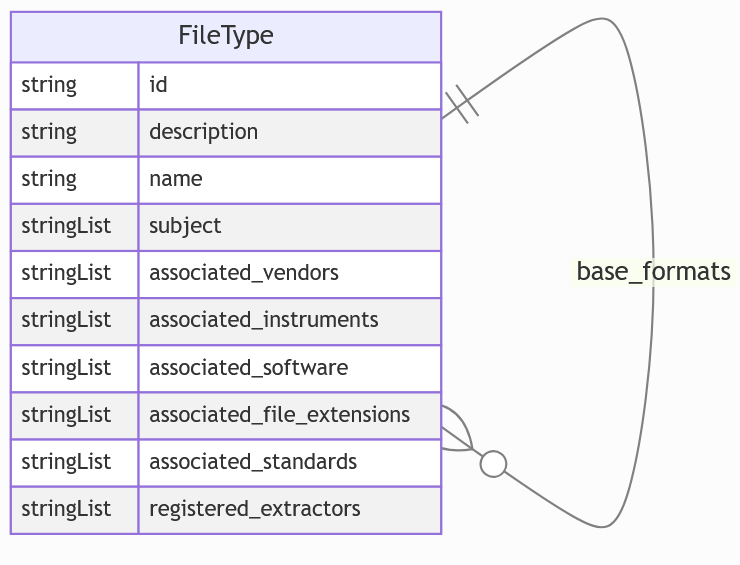}
    \vspace{-1.5em}
    \caption{The {\filetype} schema.\label{fig:ft-schema}}
    \vspace{2em} 
\end{figure}

First, the {\filetype} schema was created with the aim of allowing for the collation and curation of metadata associated with each file type. This metadata includes the relevant scientific domains, associated software, instruments, vendors, standards, or generic base formats (such as \texttt{yml} or \texttt{zip}) of a given file type.
The {\filetype} schema is relatively straightforward with a flat structure as shown in \autoref{fig:ft-schema}.
The schema contains a single loop whereby a {\filetype} can refer to another {\filetype} as its parent format. As an example, both the NeXuS and NetCDF formats can refer to HDF5 as a \texttt{base\_format}.

\begin{figure*}[tb!]
    \centering
    \includegraphics[width=1.0\linewidth]{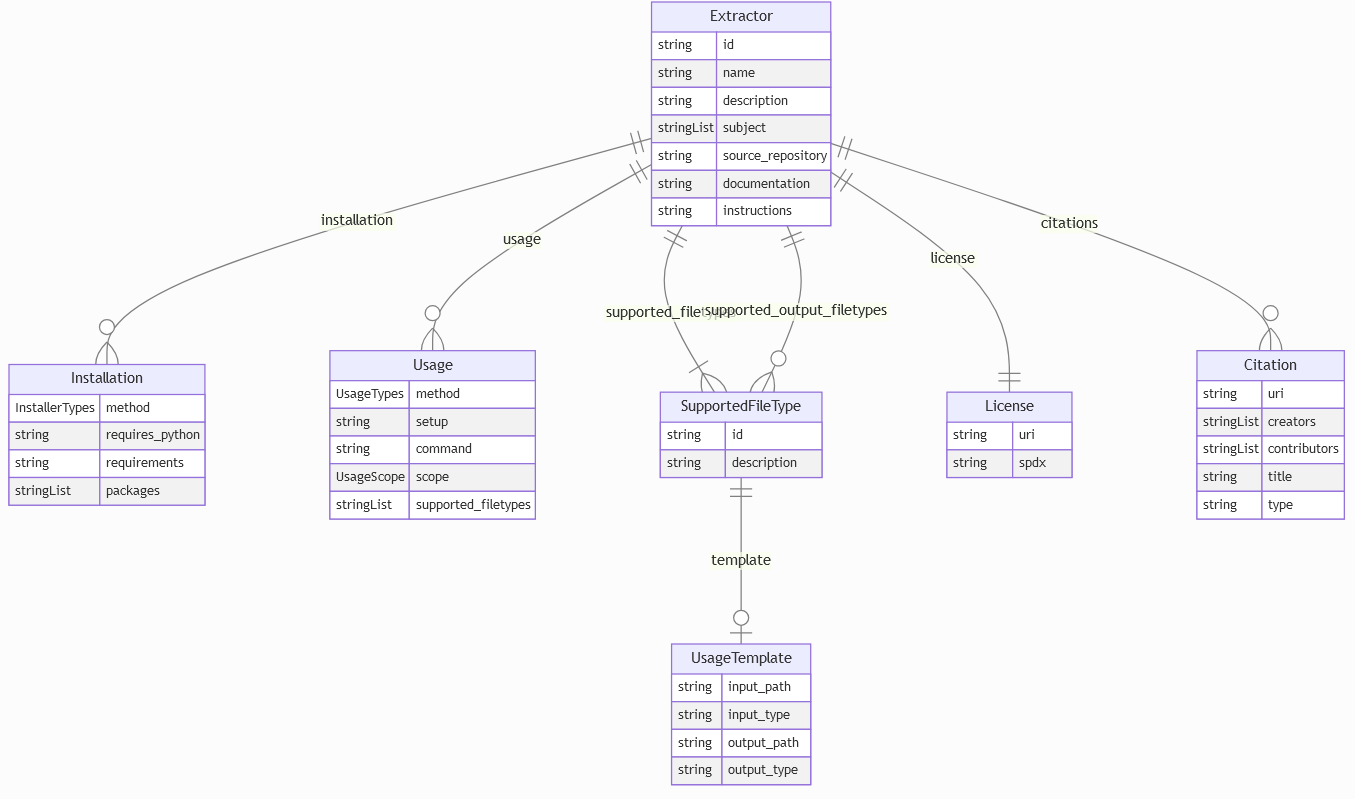}
    \vspace{-1.5em}
    \caption{The {\extractor} schema.\label{fig:ex-schema}}
\end{figure*}

The \texttt{Extractor} schema is slightly more involved, as shown in \autoref{fig:ex-schema}.
It includes submodels for describing machine-actionable installation as well as usage instructions, slots for registering which {\filetypes} it can support, as well as bibliographic and licensing information about the extractor code.
The {\extractor} schema has a built-in templating functionality, which allows the mapping of \filetype\texttt{.ID}s to appropriate switches or commands for the extractor code, actioned using the appropriate {\extractor} usage instructions. If supported by the extractor code, output {\filetype} and paths can be specified using the same templating functionality.

The two schemas were written using the programming-language-agnostic LinkML framework~\cite{Moxon2024} which enables semantic cross-walk with other schemas. Our schemas can therefore be exported to many common schema formats (e.g., JSONSchema, Pydantic, OWL) or programming language classes (Python, Java, TypeScript).
The schemas are available on GitHub,\footnote{See \href{https://github.com/datatractor/schema}{https://github.com/datatractor/schema}.} with documentation corresponding to the latest release available at \url{https://datatractor.github.io/schema}.

\subsection{Metadata registry: \texttt{yard}}
The registry is a place where the developers of extractor codes and interested users can submit, publish, search and retrieve the definitions of {\extractors} and associated \filetypes. The \texttt{yard} is the default registry for the {\datatractor} project, available at \url{https://yard.datatractor.org}. In principle, the {\datatractor} schemas can be used to populate any third party or private registry, which can be then passed to \texttt{beam} (or a similar utility) instead of the base URL of the \texttt{yard}.

Developers of extractor codes wishing to announce their extractor, or potential contributors to the \texttt{yard} are encouraged to reach out and contribute on GitHub.
To facilitate the process, we have provided Contributing Guidelines.\footnote{See \url{https://github.com/datatractor/yard}}
In short, each new software package should be submitted in a pull request to the \texttt{yard} repository and include the {\extractor} definition (see Fig.~\ref{fig:ex-yaml}) and any new supported \filetypes. Once you submit the pull request, your definitions will be validated automatically against the {\datatractor} schemas. We strongly encourage inclusion of example files for new \filetypes. The pull request will then be reviewed by one of the registry maintainers (currently the authors); after a successful review, it will be merged into the \texttt{main} branch and deployed after a new release is made. We pledge to follow the Contributor Covenant Code of Conduct.~\cite{Ehmke2014}

The \texttt{yard} can be accessed in a human-readable form (see Fig.~\ref{fig:biologic-mpr}) using the above link, where the {\filetypes} and {\extractors} can be browsed. The links from {\filetypes} to {\extractors} that support them (and vice versa) are rendered, as well as other metadata information, including the code license, links to the code repository and documentation, or references.

The machine-actionable part of the \texttt{yard} registry is the application programming interface (API), deployed at \url{https://yard.datatractor.org/api}, which is compliant with OAS 3.1.0~\cite{miller_openapi_2021} and is served using FastAPI.~\cite{Sebastian2024} Releases of the \texttt{yard} registry are automatically built using Docker, pushed onto Docker Hub, and deployed on a server at Johns Hopkins University.

\begin{figure}[tb!]
\footnotesize{
\begin{verbatim}
id: >-
  galvani
name: >-
  Galvani
description: >-
  Read proprietary file formats from 
  electrochemical test stations [...]
supported_filetypes:
  - id: biologic-mpr
  - id: biologic-mpt
license:
  spdx: GPL-3.0-only
subject:
  - electrochemistry
  - voltammetry
  - impedance spectropscopy
citations:
  - uri: https://github.com/echemdata/galvani
    creators:
      - C. Kerr
    title: galvani github repository
    type: software
source_repository: >-
  https://github.com/echemdata/galvani
usage:
  - method: python
    setup: galvani
    command: >-
      galvani.BioLogic.MPRfile({{ input_path }}).data
    supported_filetypes:
      - biologic-mpr
  [...]
installation:
  - method: pip
    packages:
      - galvani ~= 0.4
    requires_python: '>=3.6'
\end{verbatim}}
\vspace{-1.5em}
\caption{An abridged {\extractor} definition for \emph{galvani}.}\label{fig:ex-yaml}
\vspace{-1em}
\end{figure}

\subsection{Reference implementation: \texttt{beam}}

The final aspect of {\datatractor} addresses the machine-actionable usage of extractors that have been annotated via the above schemas.
In essence, this would comprise a piece of software that can automatically follow the installation instructions for the extractor code, generate the appropriate code or command-line incantation to run the extractor code on a given file, and then return the output to the user. \datatractor~\texttt{beam} is an open source reference implementation for this procedure.

For example, consider the extraction of data from an MPR file (\texttt{example.mpr}), with an associated and known {\filetype} of \texttt{biologic-mpr}, which has been generated by a BioLogic potentiostat in an electrochemical cycling experiment.
This is a proprietary, binary file format that is prone to changes across the underlying EC-Lab software releases.
The format has been reverse-engineered by multiple non-commercial open source libraries~\cite{Kraus2022a,Kerr2017,Kerr2024} to enable more complex and automated use cases than the vendor software currently allows.~\cite{Kraus2024}

With \texttt{beam}, this extraction can be done either via the command-line:

\begin{verbatim}
$ datatractor beam biologic-mpr example.mpr
Wrote output to PosixPath('example.out').
\end{verbatim}

\noindent
or in Python code directly:

\begin{verbatim}
from datatractor_beam import extract
obj = extract("example.mpr", "biologic-mpr")
\end{verbatim}


\begin{figure}[t!]
    \centering
    \includegraphics[width=1.0\linewidth]{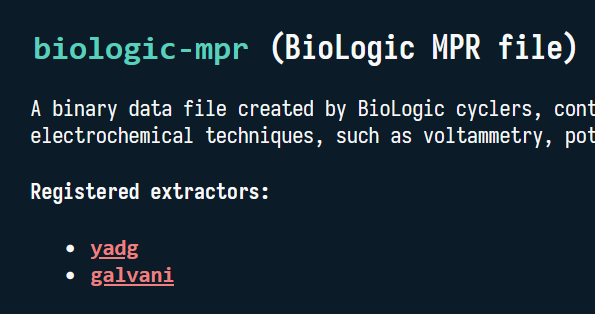}
    \vspace{-1.5em}
    \caption{A screenshot of the \texttt{biologic-mpr} {\filetype} page from \texttt{yard}, see \url{https://yard.datatractor.org/filetypes/biologic-mpr}.\label{fig:biologic-mpr}}
\end{figure}

As shown in Fig.~\ref{fig:biologic-mpr}, for the \texttt{biologic-mpr} file type, the {\datatractor} \texttt{yard} contains two matching \extractors: the \emph{yadg} library~\cite{Kraus2022a} and the \emph{galvani} package.~\cite{Kerr2017, Kerr2024} This can also be confirmed using the following command line invocation: 


\begin{verbatim}
$ datatractor probe biologic-mpr
['yadg', 'galvani']
\end{verbatim}

\noindent
When \datatractor \texttt{beam} is executed, it will fetch this information from the \texttt{yard}, and then look into the two {\extractor} definitions for a set of appropriate usage instructions that match the specified execution method (command-line or Python) and are compatible with the required \filetype. 

Currently, for usage via Python, \texttt{beam} will install and use \emph{galvani}, according to the instructions in Fig.~\ref{fig:ex-yaml}, as it is alphabetically the first {\extractor} matching the \texttt{biologic-mpr} \filetype. By default the installation is carried out in an isolated Python virtual environment for each \extractor. As \emph{galvani} does not have a command-line interface, it must be used via the Python interface. Therefore, if a command-line execution method is requested by the user, \emph{yadg} will be installed instead. Then, the \emph{yadg}-specific instructions for extracting data from the \texttt{biologic-mpr} format using the command line will be executed. In general, if the \texttt{extract()} function of \texttt{beam} is used from Python, an in-memory object will be returned (in the case of \emph{galvani} a \texttt{pandas.DataFrame}); when executing \texttt{datatractor}~\texttt{beam} from the command-line, an output file will be written instead (in the case of \emph{yadg} a NetCDF file located at \texttt{example.out}).

Much like many of the extractor codes themselves, \texttt{beam} is written in Python. It can be installed from the GitHub repository via \texttt{pip}.~\footnote{See the project README for installation instructions, \url{https://github.com/datatractor/beam?tab=readme-ov-file\#installation}.}
Without any external dependencies, \texttt{beam} uses the data present in the \texttt{yard} metadata registry to discover which {\extractors} are known for a given \filetype. Alternatively, \texttt{beam} can be pointed to a URL containing any compatible forked or modified version of the \texttt{yard} repository. Finally, \texttt{beam} can be used locally, with a local definition of an {\extractor}, by passing it as a dictionary to the \texttt{extract()} function.

With a single invocation, \texttt{beam} can then download any registered extractors and install each one in an isolated virtual environment, alleviating the common pain points with incompatible dependencies.
Finally, as in the example above, \texttt{beam} can execute the extractor code on the given file, and will return either a file converted into another format (typically a more generic format, such as JSON or HDF5), or a Python object directly.

\section{Use cases}

\subsection{Downstream use cases}

The most likely use case of the aforementioned machinery is within user-facing data management software, such as \emph{datalab}.~\cite{Evans2024}
\emph{datalab} is an open source data management platform for  materials chemistry and beyond, which is intended to be deployed at the level of an individual research group or department and can be used to track the relationships between samples, devices and characterisation data.
In order to support the many file formats that are encountered daily in this field, \emph{datalab} has explicit first-class support for several common file types, yet it is impossible to be exhaustive.
Instead, \emph{datalab} can make use of the registered extractors in the \texttt{yard}; by installing these separately from the main platform, enabling a separation of concerns between metadata extraction and data analysis.

This approach also makes it much more straightforward to develop more advanced functionality for certain file formats, as metadata extraction is often the rate-limiting step.~\cite{Statt2023}
For instance, to implement support for gas chromatography data, all that the developers of \emph{datalab} would need to write is a mapping between the detected file format and an extractor in the \texttt{yard}, and then a mapping from the extracted gas chromatography data and the desired \emph{datalab} schema for storage.
In the reverse direction, where "native" support for a file type has been implemented specifically for \emph{datalab} (for e.g., an instrument without existing supporting software implementations), \emph{datalab} developers can annotate the functionality for this file type in such a way that it can be contributed back to the \texttt{yard} for use outside of the \emph{datalab} platform itself.

With enough traction, it would also be possible to communalize the work of defining data pipelines via isolated \extractors. Multiple platforms could collaboratively develop the next iteration of \texttt{beam} that could grow to handle specific technical use cases that can arise in particular labs, involving, say, parallelisation, asynchronous or serverless processing,~\cite{Li2022} and streaming data.~\cite{Eminizer2023}
This more modular ecosystem should lower the barrier to bespoke data management platforms written and deployed where they are needed, focusing on automation and decentralization~\cite{Skluzacek2019, Skluzacek2023} rather than relying on top-down solutions to capture the long tail of scientific data that is currently left unrecorded.
The challenge is then to ensure that the \texttt{Datatractor} ecosystem is designed in such a way to enable these use cases without providing yet another source of overhead.

We note that a library for automatic {\filetype} detection is part of the electronic lab notebook (ELN) Developers Wishlist,\footnote{See the "ELN Developers Wishlist", available at~\url{https://github.com/marda-alliance/metadata_extractors/discussions/18}.} which was collected in a roundtable discussion as part of the MaRDA extractors WG. Work on such a library as well as on standardisation of \emph{outputs} of extractor codes are open avenues for further work. 

With robust {\filetype} detection in place, the {\datatractor} approach could even find use at the level of data repositories or journals, who could use it to construct indexes over uploaded data with scientifically meaningful (as opposed to generic) metadata fields. A similar approach has recently been employed for crystallographic data on the Materials Cloud Archive data repository~\cite{Talirz2020} with their integration of OPTIMADE~\cite{Andersen2021}; user archives can be automatically parsed and served with an OPTIMADE API,~\cite{Evans2021,Eimre2024} providing a standardized programmatic search interface with robust property definitions, massively improving discoverability since such data can now appear in the results of federated searches encompassing the entire OPTIMADE ecosystem.~\cite{Evans2024a}

\subsection{Upstream use cases}
The integration of high-quality third-party extractor codes in the {\datatractor} \texttt{yard} is what motivated the authors to develop this project in the first place. A list of several extractor codes we have identified is available in the GitHub discussion page of the project.\footnote{See "We Want You in Yard", available at \url{https://github.com/orgs/datatractor/discussions/2}.} Initiatives such as these also facilitate discussion about best practices: e.g., the \emph{yadg} project recently adopted the \texttt{original\_metadata} conventions used by \emph{RosettaSciIO}, thanks to the discussions sparking from this initiative.

The benefits of such integration for upstream code developers would be increased discoverability of their tools via the {\datatractor} \texttt{yard}, as well as the possibility of automated testing of {\extractor} definitions using the supplied example files of known \filetypes, especially as these examples shift to cover newer versions of vendor software over time. This enhanced visibility can allow communities of practice nucleate around particular techniques. Further self-reinforcing benefits would accrue following downstream integration of the \datatractor project.

Finally, the registry could find use in instrument procurement, where open source library support for a given file type can be a determining factor in choosing a given vendor.~\cite{Bainglass2024}
This could also encourage vendors to improve their own support for data extraction of the file types they produce.

\section{Conclusions and Outlook}
In the current work, we have proposed a registry of data extractors or format converters, intended to aid the discoverability of such tools and avoid code duplication. The registry is based on lightweight semantic schemas, which contain licensing, attribution, installation and usage instructions for such extractor codes. The registry can be used both by human users in order to find tools for extraction of their data, or by machines to perform said extraction in an automated fashion. A reference implementation using entries in this registry has also been developed. Crucially, the tools have been developed to work in a distributed fashion, allowing interested parties to spin up their own registry with minimal effort.

Some key challenges remain. First, while we developed a schema for the definition of file types, matching individual files to such file type definitions is an important but missing feature. Such a library should consider multiple hints identifying file types, including common file name extensions, magic bytes or phrases, and MIME types; the task is further complicated by the hierarchical nature of file types (a NetCDF file is a HDF5 file which is a binary file; a JSON-LD file is a JSON file which is a text file).

Secondly, while the supported input file types for extractor codes can be described using the \datatractor schemas, the output formats of those codes are currently not part of the schema. The extractor definitions could provide information describing the output formats, and an implementation of the \datatractor framework could then be used to automatically validate the outputs. Ideally, the output formats would be semantically annotated using an output schema.

Both of those outstanding issues require an amount of community consensus and contributions. The current version of the {\datatractor} framework provides a sufficiently developed yet simple prototype implementation, and a registry where upstream extractor codes can be incorporated, which at the same time may already be useful for integration into downstream projects. We hope the {\datatractor} \texttt{yard} will grow into a useful user-facing resource in its own right, and coupled with the integration of {\datatractor} \texttt{beam} (or other implementations) into downstream projects, we will be able to revise the \datatractor schemas to address the key remaining challenges, as a community. 

Whilst {\datatractor} is only a small technical piece of this puzzle, we believe that a comparable system will be crucial to enabling a future of decentralized, automated data management platforms (and thus laboratories) that can enable truly discoverable, accessible and navigable data in the physical sciences and beyond.~\cite{Skluzacek2019}

\section*{Data availability}

The three core parts of this initiative are developed on GitHub and made available under the terms of the MIT license at the \datatractor organisation: \url{https://github.com/datatractor}. The schemas, registry and \texttt{beam} implementation are archived on Zenodo at the DOIs \href{https://doi.org/10.5281/zenodo.13956909}{10.5281/zenodo.13956909}, \href{https://doi.org/10.5281/zenodo.13956953}{10.5281/zenodo.13956953}~and~\href{https://doi.org/10.5281/zenodo.13956975}{10.5281/zenodo.13956975}, respectively.

\section*{Competing interests}

M.E. is the founder and director of datalab industries ltd.

\section*{Acknowledgments}

The authors thank the members of the MaRDA Extractors WG for frequent discussions and motivation for this work, especially Edan Bainglass, Joshua Bocarsly and Steffen Brinckmann for contributions to the registry.

\section*{Funding}

M.L.E. thanks the BEWARE scheme of the Wallonia-Brussels Federation for funding under the European Commission's Marie Curie-Skłodowska Action (COFUND 847587); P.K. thanks the DFG for funding (Project Number 490703766). D.E. acknowledges support from NSF awards 2129051 (VariMat) and 2039380 (PARADIM), as well as Army Research Laboratory Cooperative Agreement Number W911NF-22-2-0101 (DSEMD).

\section*{Author contributions}

\textbf{M.L.E.}: Conceptualisation, Methodology, Software, Validation, Investigation, Data Curation, Writing - Original Draft, Writing - Review and Editing, Project Administration.
\textbf{G-M.R}: Writing - Review and Editing, Supervision, Funding Acquisition.
\textbf{D.E.}: Resources, Writing - Review and Editing, Project Administration, Funding Acquisition.
\textbf{P.K.}: Conceptualisation, Methodology, Software, Validation, Investigation, Data Curation, Writing - Original Draft, Writing - Review and Editing, Project Administration, Funding Acquisition.

\printbibliography
\end{document}